\documentclass[11pt,amsmath,amssymb,graphicx,nofootinbib]{revtex4} 
\usepackage{bm}
\usepackage{graphics,graphicx,calc,epsfig,pstricks,bbm} 
\usepackage{amsmath,amssymb,amsfonts} 
\newcommand{\be}{\begin{equation}}
\newcommand{\ee}{\end{equation}}
\newcommand{\bea}{\begin{eqnarray}}
\newcommand{\eea}{\end{eqnarray}}
\newcommand{\SO}{\mathrm{SO}}

\newcommand{\SU}{\mathrm{SU}}

\begin{document}
\title{Weaving commutators: beyond Fock space}
\author{Michele Arzano}
\email{m.arzano@uu.nl}
\affiliation{Dipartimento di Fisica,\\ ``Sapienza" University of Rome,\\ P.le A. Moro 2, 00185 Roma, EU }

\begin{abstract}
\begin{center}
{\bf Abstract}
\end{center}
The symmetrization postulate and the associated Bose/Fermi (anti)-commutators for field mode operators are among the pillars on which local quantum field theory lays its foundations.  They ultimately determine the structure of Fock space and are closely connected with the local properties of the fields and with the action of symmetry generators on observables and states.  We here show that the quantum field theory describing relativistic particles coupled to three dimensional Einstein gravity as a topological defect {\it must} be constructed using a deformed algebra of creation and annihilation operators.  This reflects a non-trivial group manifold structure of the classical momentum space and a modification of the Leibniz rule for the action of symmetry generators governed by Newton's constant.  We outline various arguments suggesting that, at least at the qualitative level, these three-dimensional results could also apply to real four-dimensional world thus forcing us to re-think the ordinary multiparticle structure of quantum field theory and many of the fundamental aspects connected to it.  
\end{abstract}

\maketitle

\section{Introduction}
Quantum field theory (QFT), the theoretical framework at the basis of our understanding of particle physics, lays its foundations on a set of fundamental assumptions whose ``raison d'etre" is intimately related with the existence of a fixed and highly symmetric background space-time.\\  
When gravity enters the quantum stage one is faced with a series of conceptual tensions which are the basis of the formidable challenge that the formulation of a quantum theory of geometry and matter has posed to theoretical physicists in the past eighty years \cite{Carlip:2001wq}.  The extent of this tension is dramatically evident already in the most celebrated effect in semiclassical gravity: black hole quantum radiance.  In this context a free quantum field living on a black hole background produces a steady thermal emission of quanta from the horizon.  Assuming an initial pure quantum state for the system, after a crude implementation of back-reaction, such {\it evaporation} would end with a final mixed state thus violating the basic postulate of unitarity of quantum evolution \cite{Hawking:1976ra}.\\  This blatant paradox, a quantum phenomenon which predicts a violation of one of the principles of quantum theory itself, forces us to pass under scrutiny all the tacit assumptions that enter the derivation of the effect. 
Factorization of the Hilbert space of the quantum field states described by a Fock space is essential in the characterization of field modes inside and outside the horizon (which in turn is closely related to locality and microcausality) and is at the basis of the assumption that the use of low-energy effective field theory is reliable in the derivation of such effect \cite{Mathur:2008wi}.  In this essay we will argue, {\it without making any assumptions about the nature of a yet-to-be-formulated theory of quantum gravity}, that three-dimensional semiclassical gravity in the presence of the simplest form of ``topological" back-reaction leads to the demise of the usual formulation of Fock space.  In particular, multiparticle states will no longer be constructed from (anti)-symmetrized tensor product of one-particle states but by a ``coloured" generalization of them reflecting a deformed algebra of creation and annihilation operators.  Newton's constant (Planck's mass in three dimensions) enters as a ``deformation parameter" which governs the non-Leibniz action of translation generators on the quantum multiparticle states.  Such deformation is a consequences of the non-trivial group manifold structure of the momentum space of the corresponding classical particle which is coupled to gravity as a topological defect.  Such unconventional quantization of the field modes signals a departure from several assumptions at the basis of ordinary quantum field theory from additivity of quantum charges associated to space-time symmetries to departures from locality.  
The fact that ordinary QFT is smoothly recovered once the Newton's constant/deformation parameter is set to zero suggests that these models can be regarded as a natural extension of the conventional field theoretic paradigm which might open new avenues in attacking the quantum gravity problem ``from below".

\section{Curved momentum space in ``flatland"}
As it is very well known Einstein gravity in three space-time dimensions does not admit local degrees of freedom \cite{Carlip:1998uc}.  Point particles are coupled to the theory as topological defects \cite{Deser:1983tn}.  The space-time describing a single particle will be flat everywhere except at the location of the particle where one has a conical singularity.  Indeed the length of a circular path centered at the location of the particle divided by its radius will be less than $2\pi$.  The deficit angle is $\alpha=8\pi G m$, proportional to the mass of the particle $m$ and Newton's constant $G$.  For the description of the phase space of the particle we need a characterization of its three-positions and three-momenta.  This can be achieved mapping the conical space-time into three-dimensional Minkowski space with a cylindrical boundary and a wedge ``cut-off" representing the deficit angle of the cone \cite{Matschull:1997du}.\\
In ordinary (non-gravitational) relativistic mechanics in three-dimensions Minkowski space is isomorphic as a vector space to the Lorentz algebra and the {\it extended} phase space is a {\it vector space} given by the direct product of two copies of $\mathfrak{sl}(2)$ i.e.  $\Upsilon\equiv \mathfrak{sl}(2) \times \mathfrak{sl}(2)\simeq \mathbb{R}^3 \times \mathbb{R}^3$.  Positions and momenta are parametrized by coordinates in such space i.e. ${\bf x}= \vec{x}\cdot \vec{\gamma}$ and ${\bf p}= \vec{p}\cdot \vec{\gamma}$ where $\gamma_a$ are $2 \times 2$ traceless matrices.\\  
When the particle is coupled to gravity we can map the simply connected part of the conical space-time to Minkowski space and thus coordinates will be given by a map ${\bf q}$ from the ``cone" to $\mathfrak{sl}(2)$.  To complete the embedding we need to map the local frames at each point of the cone into a fiducial inertial frame on the Minkowski side.  Thus to each point, besides ${\bf q}$, we also need to associate an element ${\bf U}\in SL(2)$ which provides the information regarding the type of Lorentz rotation needed to match the local frame to the background Minkowski frame.  The pair $({\bf q},{\bf U})$ provides an isometric embedding of the bundle of local frames on the simply connected part of the manifold into Minkowski space.  The freedom of choosing the background inertial frame is reflected in the freedom of transforming the functions ${\bf q}$ and ${\bf U}$ via a global Poincar\'e transformation $({\bf n}, {\bf U})$, respectively a translation and a Lorentz rotation.  Likewise the values of the embedding functions on the two faces of the wedge ${\bf q}_{\pm}$ and ${\bf U}_{\pm}$ should be identified via a generic Poincar\'e (gauge) transformation which we denote by $({\bf v}, {\bf P})$.\\
The three-position of the particle i.e. the location of its worldline in the auxiliary Minkowski space is given by the values of the function ${\bf q}$ on the cylindrical boundary which regularizes the singularity at the ``tip" of the cone.  Such function, which we denote by $\bar{{\bf q}}$, in principle depends on time and on an angular variable however, if we want the cylindrical boundary to look like a worldline, we must impose the additional condition that $\bar{{\bf q}}$ depend only on time \cite{Matschull:1997du} i.e. $\bar{{\bf q}}\equiv{\bf x}(t)$.
Thus the three-positions of the particle will be still given by a vector, namely the ``coordinates" of an element of $\mathfrak{sl}(2)$.  The values of the ``worldline" function $\bar{{\bf q}}$ at the left and right boundary of the wedge, $\bar{{\bf q}}_{\pm}$, are subject to the ``matching condition" $\bar{{\bf q}}_+\rightarrow {\bf P}^{-1} (\bar{{\bf q}}_- -{\bf v}) {\bf P}$.\\  Since $\bar{{\bf q}}_+=\bar{{\bf q}}_-={\bf x}(t)$ is the location of the particle, taking the derivative with respect to time of the equation above we obtain that {\it the velocity of the particle must commute with the group element ${\bf P}$} (remember that ${\bf v}$ is constant).  This implies that three-momentum vectors have to be proportional to the {\it projection} of the group element ${\bf P}\in SL(2)$ on its Lie algebra $\mathfrak{sl}(2)$, i.e. if we write ${\bf P}$ in its matrix expansion ${\bf P} = u \mathbbm{1} + 4\pi G \vec{p} \cdot \vec{\gamma}$, we discard the part of ${\bf P}$ proportional to the identity and take ${\bf p}=\vec{p} \cdot \vec{\gamma}$.  Notice that now the components of the momentum vector are {\it coordinates on a group manifold}, indeed the condition $\det {\bf P} =1$ implies that $u^2 - 16\pi^2 G^2 \vec{p}^{\, 2} = 1$,  the equation of a hyperboloid embedded in $\mathbb{R}^4$.  The phase space in the presence of topological ``gravitational backreaction" is thus $\Upsilon_G =  \mathfrak{sl}(2) \times SL(2)\simeq \mathbb{R}^3 \times SL(2)$.\\
In the following sections we will discuss the dramatic consequences that this ``structural" modification of the phase space of a relativistic particle has for the corresponding (quantum) field theory.

\section{From ``conical" particles to ``braided" commutators}
\subsection{One particle}
We turn now to the description of the field theory describing the quantization of the relativistic particle coupled to three-dimensional gravity discussed in the Introduction.  Before we do that it will be useful to make a short digression on the definition of the physical phase space and the associated mass-shell relation. In order to determine the mass shell we need to find a characterization of the mass of the particle and relate it to the notion of generalized momentum.  As we saw above the mass of the particle is proportional to the deficit angle of the conical space.  A way to measure the deficit angle is to transport a vector along a closed path around the boundary, as a result this will be rotated by the angle $\alpha= 8 \pi G m$.  Physical momenta will be thus characterized by ``holonomies" $\bar{\bf{P}}$ which represent a rotation by $\alpha= 8 \pi G m$.  Such requirement imposes the restriction 
\be
\frac{1}{2}\mathrm{Tr} (\bar{{\bf P}}^2) = \cos (4\pi Gm)\,\,\,\longrightarrow \,\,\,\,  \vec{p}^{\, 2} = - \frac{\sin^2(4\pi G m)}{16 \pi^2 G^2}\, ,
\ee
on the ``physical" holonomies giving us a ``deformed" mass-shell condition.  From a mathematical point of view such on-shell condition is equivalent to imposing that physical holonomies/momenta lie in a given {\it conjugacy class} of the Lorentz group (see \cite{Schroers:2007ey} for a pedagogical discussion). Roughly speaking if in ordinary Minkowski space the mass-shell hyperboloid describing the physical momenta of a massive particle is given by an orbit of the Lorentz group in flat momentum space, physical momenta belonging to a given conjugacy class can be seen as ``exponentiated" orbits of the group on its manifold.\\  
In analogy with ordinary Minkowski space we can consider complex functions on the momentum group manifold described above.  It turns out that when restricted to momenta belonging to a given conjugacy class, such space of functions carries a unitary irreducible representation of the semi-direct product of the momentum group manifold and the Lorentz group \cite{Koornwinder:1998xg} and thus is analogous (modulo a choice of polarization \cite{Arzano:2010jw}) to the ordinary one-particle Hilbert space for a quantum field.\\  Without loss of generality and to keep our considerations at the simplest level we now switch to Euclidean signature allowing the ``phase space" of the particle to be $\Upsilon_G =  \mathfrak{su}(2) \times SU(2)$.  In analogy with ordinary field theory we consider plane waves labelled by group elements belonging to a given conjugacy class as representatives of a one-particle ``wave function".  Momenta are now coordinates on $SU(2)$, in particular we work with ``cartesian" coordinates\footnote{For simplicity we restrict to functions on $\SO(3)\simeq\SU(2)/\mathbb{Z}_2$.}
\begin{equation}
\label{gp}
{\bf P}(\vec{p}) = p_0 \, \mathbbm{1}+ i \,  \, \frac{\vec{p}}{\kappa} \cdot \vec\sigma,
\end{equation}
where  $\kappa=(4\pi G)^{-1}$, $p_0=\sqrt{1-\frac{\vec{p}^2}{\kappa^2}}$ and $\vec{\sigma}$ are Pauli matrices.  Plane waves can be written in terms of a Lie algebra element  ${\bf x}= x^i \sigma_i \in\mathfrak{su}(2)$ as
\be
e_{\bf P}(x) = e^{\frac{i}{2\kappa} \mathrm{Tr} ({\bf x} {\bf P})}=e^{i\vec{p}\cdot\vec{x} }\, .
\ee
with $\vec{p}=\frac{\kappa}{2i} \mathrm{Tr} ({\bf P} \vec{\sigma})$.  The main effect of the group structure of momentum space is that the composition of plane waves is {\it non-abelian} indeed we can define a $\star$-product for plane waves
\be
e_{\bf P_1}(x) \star e_{\bf P_2}(x) = e^{\frac{i}{2\kappa} \mathrm{Tr} ({\bf x} {\bf P_1})} \star e^{\frac{i}{2\kappa} \mathrm{Tr} ({\bf x} {\bf P_2})} = e^{\frac{i}{2\kappa} \mathrm{Tr} ({\bf x} {\bf P_1 P_2})}\, ,
\ee
differetiating both sides of this relation one can easily obtain a non-trivial commutator for the $x$'s 
\be\label{spinsp}
[x_l, x_m] = i \kappa \epsilon_{lmn} x_n\, ,
\ee
i.e. the ``coordinates" $x_i$ can be seen as equipped with a non-commutative algebra structure.  Functions of these coordinates will inherit a non-abelian algebra structure and the corresponding field theory will be a {\it non-commutative field theory}.  Most importantly momenta will obey a non-abelian composition rule due to the non-trivial group structure  
\be
\vec{p}_{1} \oplus  \vec{p}_{2} =  p_0(\vec{p}_{2})\, \vec{p}_{1} + p_0(\vec{p}_{2})\, \vec{p}_{2} + \frac{1}{\kappa} \vec{p}_{1} \wedge \vec{p}_{2}=
 \vec{p}_{1} + \vec{p}_{2} + \frac{1}{\kappa} \vec{p}_{1} \wedge \vec{p}_{2}+ \mathcal{O}(1/\kappa^2)\, .
\ee
Since plane waves are eigenfunctions of translation generators the non-abelian composition of momenta will correspond to a non trivial action of translation generators on multiparticle states, in particular one can easily derive the following generlization of the Leibniz rule on the tensor product of two one-particle states
\be
\Delta P_a =  P_a\otimes {\bf 1} + {\bf 1}\otimes P_a + \frac{1}{\kappa} \, \epsilon_{abc} P_b \otimes P_c + \mathcal{O}(1/\kappa^2)\, .
\ee
Notice that $\frac{1}{\kappa} = 4\pi G$ can be seen as a {\it deformation parameter} and in the limit $\kappa\rightarrow \infty$ one recovers the usual action of translations as an abelian Lie algebra.  As discussed in \cite{Arzano:2007nx} such behaviour signals a departure from one of the basic postulates of quantum field theory namely the additivity of the charges associated with space-time symmetry generators which in turn is deeply connected with the locality properties of the field operators \cite{Haag:1974qh}.\\

\subsection{More particles}
So far we have seen that at the one particle level the mass-shell condition which in the ordinary case is given by orbits of the Lorentz group on the vector space $\mathbb{R}^{3,1}$ is replaced now by ``orbits" of the Lorentz group on a {\it non-linear} momentum space or, more properly, {\it conjugacy classes}.   In analogy with ordinary field theory let us label one particle states by elements of these conjugacy subspaces of $SU(2)$ and denote them by $|{\bf P}\rangle$.  As for any quantum system the space of states of a composite object is built from the tensor product of its constituents Hilbert spaces.  Since at the quantum level we are dealing with indistinguishable particles one {\it postulates} \cite{Messiah:1900zz} that $n$-particle states are constructed from (anti)-symmetrized $n$-fold tensor products of one-particle Hilbert spaces for particles with (half)-integer spin.  Focusing on the simplest case of a two-particle state one notices that naively adopting a standard symmetrization prescription (we assume for definiteness that we are dealing with a spinless particle) the candidate state $1/\sqrt{2} \left( |{\bf P_1}\rangle \otimes |{\bf P_2}\rangle + |{\bf P_2}\rangle \otimes |{\bf P_1}\rangle\right)$ {\it is not} an eigenstate of the translation operator due to the non-abelian composition of momenta (reflecting the non-trivial Leibniz action of translation generators discussed above).  This problem can be bypassed if one resorts to the following ``momentum dependent" symmetrization \cite{Arzano:2008bt, ArzKow2012}
\be
|{\bf P_1}{\bf P_2}\rangle_L \equiv 1/\sqrt{2} \left( |{\bf P_1}\rangle \otimes |{\bf P_2}\rangle + |{\bf P_1}{\bf P_2}{\bf P_1^{-1}}\rangle \otimes |{\bf P_1}\rangle\right)\,.
\ee
We used the subscript $L$ above because one could also choose
\be
|{\bf P_1}{\bf P_2}\rangle_R \equiv 1/\sqrt{2} \left( |{\bf P_1}\rangle \otimes |{\bf P_2}\rangle + |{\bf P_2}\rangle \otimes |{\bf P_2^{-1}}{\bf P_1}{\bf P_2}\rangle\right)\,.
\ee
Notice how now both two-particle states are eigenstates of the generators $P_a$ and have total on-shell momentum ${\bf P_1}{\bf P_2}\equiv \vec{p}_{1} \oplus  \vec{p}_{2}$.  In analogy with the standard case we can introduce creation and annihilation operators so that
\be
a^{\dagger}_{L,R}({\bf P_1}) a^{\dagger}_{L,R}({\bf P_2}) |0\rangle \equiv |{\bf P_1}{\bf P_2}\rangle_{L,R}\, .
\ee
The action of the Lorentz group on the kets will be given by conjugation \cite{Sasai:2007me} i.e. 
\be
\Lambda_{\bf H} \triangleright |{\bf P_i}\rangle\equiv |{\bf H^{-1}}{\bf P_i} {\bf H}\rangle\, .
\ee 
It is straightforward to check that both ``left" and ``right" symmetrizations above are {\it covariant} under such transformations \cite{ArzKow2012}.  Moreover ``L" and ``R"-symmetrized states are connected by a Lorentz transformation
\be
(\Lambda_{{\bf P_1}}\otimes \Lambda_{{\bf P_2}}) \triangleright |{\bf P_1}{\bf P_2}\rangle_L =  |{\bf P_1}{\bf P_2}\rangle_R\,.
\ee
In order to determine the algebra satisfied by such operators we start by noticing the useful relation 
\be
 (\Lambda_{{\bf P_2}^{-1}}\circ \Lambda_{{\bf P_1}})\otimes \mathbf{1} \triangleright |{\bf P_1}{\bf P_2}\rangle_L =  |{\bf P_2} {\bf P_1}\rangle_L\,.
\ee
Defining $\mathcal{R}^{-1}_L({\bf P_1}, {\bf P_2}) \equiv  (\Lambda_{{\bf P_2}^{-1}}\circ \Lambda_{{\bf P_1}})\otimes \mathbf{1}$ we can then write the following {\it braided} commutators
\bea
a^{\dagger}_{L}({\bf P_1}) a^{\dagger}_{L}({\bf P_2}) - \mathcal{R}^{-1}_L({\bf P_1}, {\bf P_2}) a^{\dagger}_{L}({\bf P_2}) a^{\dagger}_{L}({\bf P_1})   & = & 0 \\
a_{L}({\bf P_1}) a_{L}({\bf P_2}) - \mathcal{R}_L({\bf P_1}, {\bf P_2}) a_{L}({\bf P_2}) a_{L}({\bf P_1})  & = & 0 \,.
\eea
One can proceed in an analogous way for the {\it right} operators to find similar commutation relations with $\mathcal{R}^{-1}_L({\bf P_1}, {\bf P_2})$ replaced by $\mathcal{R}^{-1}_R ({\bf P_1}, {\bf P_2}) \equiv \mathbf{1}\otimes (\Lambda_{{\bf P_1}}\circ\Lambda_{{\bf P_2^{-1}}})$.  
The cross-commutators between $a({\bf P})$ and $a^{\dagger}({\bf P})$ will be similarly \cite{ArzKow2012} given by 
\be
a_L({\bf P_1}) a^{\dagger}_{L}({\bf P_2}) - \mathcal{\hat{R}}_{L}({\bf P_2})a^{\dagger}_{L}({\bf P_2}) a_L({\bf P_1}) =\delta({\bf P_1^{-1}}{\bf P_2})
\ee
where $\mathcal{\hat{R}}_{L}({\bf P_2})= \mathbf{1}\otimes \Lambda_{{\bf P_2}}$ and $\delta({\bf P_1^{-1}}{\bf P_2})$ is the Dirac delta function on the group \cite{VileK}.  One can proceed in an analogous way for the ``R" operators and obtain
\be
a_R({\bf P_1}) a^{\dagger}_{R}({\bf P_2}) - \mathcal{\hat{R}}_{R}({\bf P_2})a^{\dagger}_{R}({\bf P_2}) a_R({\bf P_1}) =\delta({\bf P_1^{-1}}{\bf P_2})
\ee
where now $\mathcal{\hat{R}}_{R}({\bf P_1})= \Lambda_{{\bf P_1}} \otimes \mathbf{1}$.\\
We arrived to a modification of the usual algebra of creation and annihilation operators which is quite suggestive.  It is reminiscent of the algebra of q-deformed oscillators or ``quons" \cite{Greenberg91} but with ``colour"-dependent q-factors given by $\mathcal{R}_{L,R}({\bf P_1}, {\bf P_2})$ and $\mathcal{\hat{R}}_{L,R}({\bf P})$.  We leave it open to speculation whether such deformed commutators can be interpreted as the quantum counterpart of the braiding of the worldlines of classical point-defects.

\section{Discussion}
The familiar form of the algebra of creation and annihilation operators that we are accustomed to from quantum field theory textbooks is intimately related to the quantization condition one imposes on fields and their conjugate momenta.  The latter is {\it assumed} on the basis of the analogy with ordinary quantum mechanical commutators between position and momenta of a non-relativistic particle.  
The results we presented show that Einsten gravity in three space-time dimensions {\it clearly} indicate a possible relaxation of such assumption and a departure from the basic structures underlying our familiar formulation of local quantum field theory.
The most immediate consequence of the deformed algebra of oscillators, as we showed above, is that the Fock space of the theory loses its simple structure in terms of (anti)-symmetrized tensor products of given one-particle states.  It has been suggested \cite{AmelinoCamelia:2007zzb} that these types of departures from ordinary Fock space might reflect a new kind of uncertainty on the measurement of momenta of multiparticle states namely that measuring the total momentum of a system precludes complete knowledge of the total momenta of its components and vice-versa.  Besides this observation what is evident now is that due to the ``braided" nature of the multiparticle states the question of decoupling of the low energy degrees of freedom form the high energy ones must be handled with care.  This could suggest a weak link in the assumptions at the basis of the derivation leading to the information paradox, namely the use of low energy effective field theory in the presence of backreaction.\\
Another key aspect that is put at stake in these models is locality.  In the discussion above we briefly touched upon the the fact that the Leibniz action of symmetry generators on quantum states is deeply connected with the local properties of the fields.  It turns out that allowing a non-trivial geometry for the momentum space of a classical particle has been subject to recent investigations in the context of the ``relative locality" paradigm \cite{AmelinoCamelia:2011bm}.  The phase space of a particle coupled to three dimensional gravity can indeed be seen as an example of a relative locality theory \cite{AmelinoCameliaArz2012}.  The conceptual breakthrough of such models lies in the observer-dependent notion of crossing of particle worldlines.  The far reaching implications of this new feature have been widely discussed in the literature \cite{AmelinoCamelia:2011nt, AmelinoCamelia:2011pe}.  In the perspective of our discussion it will be useful to investigate the behaviour of field operators constructed via the deformed operators above in order to check whether ``classical" relative locality translates at the quantum level into departures from the ordinary local field paradigm.\\
Of course all the discussion so far is very specific to three dimensional gravity and its topological nature.  What about the more realistic four-dimensional world?  Obviously in four space-time dimensions Einstein's gravity {\it is not} a topological theory and thus in general similar arguments would not hold.  Surprisingly though there exist suggestive results on Planckian scattering in quantum gravity that appear to hint in the right direction. Early work by 't Hooft \cite{'tHooft:1987rb} and by Verlinde and Verlinde in the early 90's \cite{Verlinde:1991iu} showed that forward scattering at Planckian center of mass energies in 3+1 quantum gravity can be treated semiclassically and gravity splits in a weakly coupled sector and a strongly coupled sector whose quantum dynamics can be described by a topological field theory.  Could we be dealing with a similar state of affairs also in this four dimensional regime?  As of today the question remains open.\\
The recent framework of {\it piecewise flat} gravity in $3+1$ dimensions \cite{'tHooft:2008kk} proposed as a model for gravity which displays only a finite number of degrees of freedom per compact regions of space-time could also provide a bridge to the real four dimensional world.  Indeed this model is based on a straightforward extension of the picture of a system of particles described as defects which is found in three dimensional gravity.  To our knowledge nobody has attempted a study of the phase space of these particles/strings in the same spirit of \cite{Matschull:1997du}.  It would be not surprising if one would end up finding non-trivial structures analogous to the ones we discussed in this essay.\\
Finally, following the ``relative locality" framework mentioned above one could argue that a curved momentum space is just a feature of a regime of four dimensional {\it quantum gravity} in which the Planck length is negligible while the Planck mass remains finite \cite{AmelinoCamelia:2011pe}. This formally means that both quantum and local gravitational effects become negligible, while their ratio remains finite and governs the non-trivial geometry of momentum space.  If this assumptions are correct then our arguments would qualitatively hold true in four dimensions and they would indicate that ``first order" quantum gravity corrections to local QFT would be exactly the kind described above.\\
In our opinion and in the light of the observations above, large part of the conceptual apparatus of local QFT is ripe for re-thinking and the three dimensional world is there to point us the way to go beyond the various assumptions that lie their roots in the very structure of Minkowski space.  What we find remarkable is that the simple combination of {\it ordinary} classical gravity and quantum theory (via a topological coupling), without any reference to a specific ``quantum gravity" model, suggests that departures from local QFT become quite natural when gravity enters the game.  This suggests that the ``humble" framework of semiclassical gravity has still a lot to teach us on various puzzling aspects of the marriage between gravity and the quantum world.

\begin{acknowledgments}
I would like to thank J. Kowalski-Glikman and V. De Carolis for discussions.  
This work is supported by EU Marie Curie Actions and in part by a grant from the John Templeton Foundation. 
\end{acknowledgments}


\begin{thebibliography}{99}

\bibitem{Carlip:2001wq} 
  S.~Carlip,
  Rept.\ Prog.\ Phys.\  {\bf 64}, 885 (2001)
  [gr-qc/0108040].

\bibitem{Hawking:1976ra} 
  S.~W.~Hawking,
  Phys.\ Rev.\ D {\bf 14}, 2460 (1976).
  
\bibitem{Mathur:2008wi} 
  S.~D.~Mathur,
  Lect.\ Notes Phys.\  {\bf 769}, 3 (2009)
  [arXiv:0803.2030 [hep-th]].

\bibitem{Carlip:1998uc} 
  S.~Carlip,
  Cambridge, UK: Univ. Pr. (1998) 276 p
  
\bibitem{Deser:1983tn}
  S.~Deser, R.~Jackiw, G.~'t Hooft,
  Annals Phys.\  {\bf 152}, 220 (1984).  

\bibitem{Matschull:1997du}
  H.~J.~Matschull, M.~Welling,
  Class.\ Quant.\ Grav.\  {\bf 15}, 2981-3030 (1998).
  [gr-qc/9708054].  
  
\bibitem{Schroers:2007ey} 
  B.~J.~Schroers,
  PoS QG {\bf -PH}, 035 (2007)
  [arXiv:0710.5844 [gr-qc]].

\bibitem{Koornwinder:1998xg} 
  T.~H.~Koornwinder, F.~A.~Bais and N.~M.~Muller,
  Comm.\ Math.\ Phys.\  {\bf 198}, 157 (1998)
  [q-alg/9712042].

\bibitem{Arzano:2010jw} 
  M.~Arzano,
  Phys.\ Rev.\ D {\bf 83}, 025025 (2011)
  [arXiv:1009.1097 [hep-th]].
  
\bibitem{Arzano:2007nx} 
  M.~Arzano,
  Phys.\ Rev.\ D {\bf 77}, 025013 (2008)
  [arXiv:0710.1083 [hep-th]].
  
\bibitem{Haag:1974qh} 
  R.~Haag, J.~T.~Lopuszanski and M.~Sohnius,
  Nucl.\ Phys.\ B {\bf 88}, 257 (1975).
  
\bibitem{Messiah:1900zz} 
  A.~M.~L.~Messiah and O.~W.~Greenberg,
  Phys.\ Rev.\  {\bf 136}, B248 (1964).
  
\bibitem{Arzano:2008bt} 
  M.~Arzano and D.~Benedetti,
  Int.\ J.\ Mod.\ Phys.\ A {\bf 24}, 4623 (2009)
  [arXiv:0809.0889 [hep-th]].

\bibitem{ArzKow2012} 
  M.~Arzano and J.~Kowalski-Glikman,  to appear.

\bibitem{Sasai:2007me} 
  Y.~Sasai and N.~Sasakura,
  Prog.\ Theor.\ Phys.\  {\bf 118}, 785 (2007)
  [arXiv:0704.0822 [hep-th]].

\bibitem{VileK}
N.J.~Vilenkin and A.U.~Klimyk,
Kluwer, 1991, 1993.

\bibitem{Greenberg91}
O. W. Greenberg, 
Phys. Rev. D {\bf 43}, 4111-4120 (1991)

\bibitem{AmelinoCamelia:2007zzb} 
  G.~Amelino-Camelia, A.~Marciano and M.~Arzano,
  In *Di Domenico, A. (ed.): Handbook of neutral kaon interferometry at a Phi-factory* 155-186
  
\bibitem{AmelinoCamelia:2011bm} 
  G.~Amelino-Camelia, L.~Freidel, J.~Kowalski-Glikman and L.~Smolin,
  Phys.\ Rev.\ D {\bf 84}, 084010 (2011)
  [arXiv:1101.0931 [hep-th]].
  
\bibitem{AmelinoCameliaArz2012} 
G.~Amelino-Camelia, M. Arzano, S. Bianco and R. Buonocore, to appear; 
 J.~Kowalski-Glikman and  T.~Trzesniewski, to appear. 
 
\bibitem{AmelinoCamelia:2011nt} 
  G.~Amelino-Camelia, M.~Arzano, J.~Kowalski-Glikman, G.~Rosati and G.~Trevisan,
  Class.\ Quant.\ Grav.\  {\bf 29}, 075007 (2012)
  [arXiv:1107.1724 [hep-th]].
  
\bibitem{AmelinoCamelia:2011pe} 
  G.~Amelino-Camelia, L.~Freidel, J.~Kowalski-Glikman and L.~Smolin,
  Gen.\ Rel.\ Grav.\  {\bf 43}, 2547 (2011)
  [Int.\ J.\ Mod.\ Phys.\ D {\bf 20}, 2867 (2011)]
  [arXiv:1106.0313 [hep-th]].
  
\bibitem{'tHooft:1987rb} 
  G.~'t Hooft,
  Phys.\ Lett.\ B {\bf 198}, 61 (1987).
  
\bibitem{Verlinde:1991iu} 
  H.~L.~Verlinde and E.~P.~Verlinde,
  Nucl.\ Phys.\ B {\bf 371}, 246 (1992)
  [hep-th/9110017].
  
\bibitem{'tHooft:2008kk} 
  G.~'t Hooft,
  Found.\ Phys.\  {\bf 38}, 733 (2008)
  [arXiv:0804.0328 [gr-qc]].
\end{thebibliography}
\end{document}